\documentclass[aps,pre,twocolumn]{revtex4-1}
\usepackage{amsmath}
\usepackage{verbatim}
\usepackage{amsfonts}
\usepackage{amssymb}
\usepackage{graphicx}
\usepackage[colorlinks,linkcolor=blue,anchorcolor=blue,citecolor=blue,urlcolor=black]%
{hyperref}
\usepackage{mathrsfs}
\usepackage{dcolumn}
\usepackage{bm}
\usepackage{epsfig}
\usepackage{calrsfs}
\usepackage{calligra}
\usepackage[version=3]{mhchem}%
\DeclareMathAlphabet{\mathcal}{T1}{calligra}{m}{n}
\DeclareFontFamily{OT1}{pzc}{}
\DeclareFontShape{OT1}{pzc}{m}{it}{<-> s * [1.10] pzcmi7t}{}
\DeclareMathAlphabet{\mathpzc}{OT1}{pzc}{m}{it}
\def\be{\begin{equation}}
\def\ee{\end{equation}}
\begin{document}
\title{Counting statistics of chaotic resonances at optical frequencies:  \newline
 theory and experiments}
\author{Domenico Lippolis$^{1}$}
\email{domenico@ujs.edu.cn}
\author{Li Wang$^{2}$}
\author{Yun-Feng Xiao$^{2,3}$}
\altaffiliation{URL: www.phy.pku.edu.cn/$\sim$yfxiao/}

\affiliation{$^{1}$Institute for Applied Systems Analysis, Jiangsu University, Zhenjiang 212013, China}
\affiliation{$^{2}$State Key Laboratory for Mesoscopic Physics and School of Physics,
Peking University; Collaborative Innovation Center of Quantum Matter, Beijing 100871, China}
\affiliation{$^{3}$Collaborative Innovation Center of Extreme Optics Taiyuan 030006, Shanxi, P. R. China}

\date{\today}

\begin{abstract}
A deformed dielectric microcavity is used as an experimental platform for the analysis of the statistics of
chaotic resonances, in the perspective of testing fractal Weyl laws at optical frequencies.
In order to surmount the difficulties that arise from reading strongly overlapping spectra, we exploit
the mixed nature of the phase space at hand, and only count the high-Q whispering-gallery modes (WGMs) directly. That enables us to draw statistical information on the more lossy chaotic resonances, coupled to the high-Q regular modes via dynamical tunneling. Three different models [classical, Random-Matrix-Theory (RMT) based, semiclassical] to interpret the experimental data are discussed. On the basis of least-squares analysis, theoretical estimates of Ehrenfest time, and independent measurements, we find that a semiclassically modified RMT-based expression best describes the experiment in all its realizations, particularly when the resonator is coupled to visible light, while RMT alone still works quite well in the infrared.
In this work we reexamine and substantially extend the results of a short paper published earlier
[L. Wang, D. Lippolis, Z.-Y. Li, X.-F. Jiang, Q. Gong, and Y.-F. Xiao, Phys. Rev. E \textbf{93}, 040201(R) (2016)].

\end{abstract}

\maketitle

\section{Introduction}

Confinement and manipulation of photons using whispering gallery mode (WGM)
microcavities \cite{McCall92,microcavity03,Matsko06} have triggered intense
research due to their unique features, such as the long photon lifetime and
strong field confinement. By breaking the rotational symmetry of the WGM
microcavities \cite{CaoWier}, it was recently found that the deformed
microcavities not only gain directionality, highly desirable for microlasers
and other photonics applications
\cite{Benson11,SunaHara,broadening,park09,nature97,gmachl98,jan08,susumu10,song10,song12,XF_dfmc},
 but also serve as dynamical billiards for experimentally testing the systems
with a mixed phase space, from which one can study classical and quantum chaos
\cite{Wisersig06,jonathan09}. In particular, prominent phenomena were so far
demonstrated experimentally in the optical microcavity system, \textit{e.g.},
dynamical tunneling \cite{susumu10,susumu11,Yang_dyntun,xiao13}, dynamical
localization \cite{Podolskiy,Fang}, scarring
\cite{KorScar,YaleScar,GmachlScar}, turnstile transport \cite{shim08}, and
avoided resonance crossings \cite{KorCross}.

The study of quasibound states (resonances), of importance to
understand the mechanisms of chaotic scattering~\cite{BR,Gasp89,Anlage,squid}, is not
so often performed on dielectric microcavities~\cite{WierMain}. That is due to both
experimental and theoretical challenges. On the experimental
side, chaotic resonances are often very lossy, and tend to overlap
in the spectrum, making recognition problematic~\cite{Zwor_pre}.
From a theoretical standpoint, the observations may lend themselves  to
multiple interpretations~\cite{Nonnenmacher,Schonwetter,Schenck},
 due to the \textit{partial} openness of the system, which
makes the wave-ray correspondence highly
nontrivial~\cite{APT13}.

\begin{figure}[tbh!]
\centerline{\includegraphics[width=8cm]{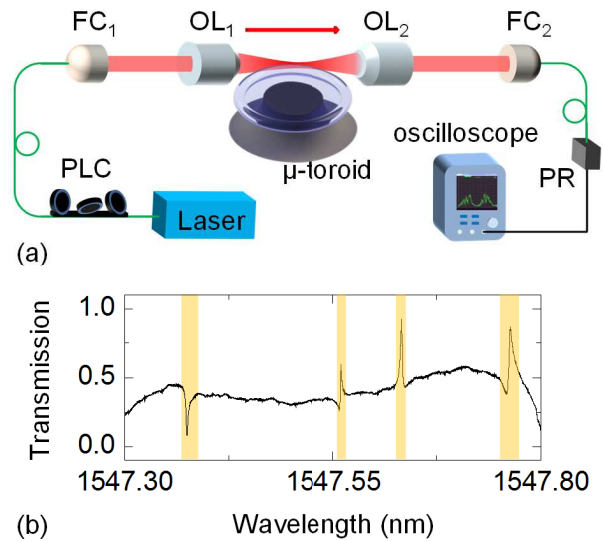}}
\caption{(color online) (a) Schematic representation of the free-space coupled cavity
system: the cavity field is excited by visible or infrared laser, while
the transmitted signal is detected by an oscilloscope. Key: PLC = polarization
controller; FC = fiber coupler;  OL = optical lens; PR = photon receiver.
 (b) A typical transmission spectrum with the high-$Q$ regular
modes highlighted.}%
\label{figureone}%
\end{figure}

 In the present work we propose a solution to the above problems, by employing  a  silica-made microcavity~(Fig.~\ref{figureone}, also see ref.~\cite{LiRC16}), which is
approximately two-dimensional, has a deformed circular boundary, and
is placed on the top of a silicon-made, fully-absorbing pedestal.

The ray dynamics inside the resonator, which can be regarded as
a leaky billiard, is mixed:
while the chaotic
dynamics mostly dwells in the central region, the regular (quasiperiodic) rays
closely follow the boundary and thus live in the outer toroid. Although there is no physical boundary
dividing these two regions of the cavity, regular and chaotic dynamics are well separated in
 the phase space (Fig.~\ref{SOS}) by a KAM boundary~\cite{Lazutkin}, so that no classical trajectory can cross between the two.
 Quantum mechanically, however, a wave localized on one region of the phase space may tunnel into the
 other~\cite{davis81}, so that, generally, overlapping chaotic resonances are coupled to sharp,
 non-overlapping
 regular (WGM) ones~\cite{HackNock}, and counting the latter from the transmission spectrum can help us draw information on the statistics of the former. That is the basic strategy we adopt to avoid the
 problem of reading overlapping spectra.

Moreover, the microresonator used here is fabricated on the top of a silicon pillar of smaller radius, which fully absorbs
virtually every ray that travels directly above it. Consequently, the system acquires a full opening, and
the present experiments may be used to validate existing predictions for the statistics of chaotic resonances.

With these premises, the
experiments performed here are aimed at estimating the number of
\textit{chaotic} resonances from the sole observation of \textit{regular} ones, mostly WGMs.
A thorough analysis is also presented, where we test three different models against the
experimental data: $i)$
a classical prediction, solely based
on ray dynamics, $ii)$ a known expression~\cite{ZycSom} obtained from the
truncation of random matrices , and $iii)$ a semiclassical correction~\cite{SchomTwor} to $ii)$,
which depends on the Lyapunov exponent of the chaotic dynamics, and therefore on
system-specific properties. Theory, methods, experimental conditions, and statistical analysis
are explained in full detail.

The method we introduce is intended to set the stage for more general investigations of
chaotic scattering phenomena in open systems, beginning with a
test at optical frequencies of  the fractal Weyl law for the scaling of the number of resonances
with the energy~\cite{LSZ}.
At present, the scaling exponent in
this prediction is also believed to depend
upon the cutoff chosen for the linewidth~\cite{NonZwor05,WierMain}
of the resonances counted, and thus on the range of decay times
of the corresponding chaotic states.
With that in mind, an important aspect of the present analysis
is that to estimate the typical decay time that the experiments
are sensitive to. A comparison 
of the measured maximum escape rates of  the chaotic states
with the estimated Ehrenfest time of quantum-to-classical correspondence~\cite{SchomJacq}
provides useful information in that respect.

The paper is organized as follows. Section~\ref{Theory} contains the theoretical model, with
the equations that couple chaotic- to regular modes, obtained with two different but equivalent
approaches (Secs.~\ref{Anth} and~\ref{fanoth}, respectively). The key relation between number of
chaotic modes and probability of excitation of one regular mode is derived in Sec.~\ref{keysec}.
The statistics of chaotic modes is treated in Sec.~\ref{statchastat} with different models, that depend
on the timescales involved. In Sec.~\ref{absorber}, we place an absorber at the center of the cavity to
obtain a full opening. Varying the size of the absorber affects the mean dwell time of the
chaotic rays from the cavity. A theoretical study of  the statistics of resonances and
the number of regular modes excited as a function of the radius of the absorber is presented
in  Sec.~\ref{WGMexct}, while in Sec.~\ref{transch}
we numerically investigate the time scale of transient chaos
versus escape to the absorber in the ray dynamics.
The experimental apparatus
is described in Sec.~\ref{expset},  together with the techniques employed to
perform measurements of the transmission spectra, and
numerical studies of the propagation of both chaotic modes and
WGMs inside the cavity.
Section~\ref{results}  contains the experimental results and their
statistical analysis: regular modes are counted in various
experimental conditions, and the three different models are validated
against the data (Secs.~\ref{class_stat} and~\ref{RMTSC}).
We discuss the proportionality between the
probability of excitation of a single regular mode and the number
of excited regular modes
in Sec.~\ref{linewidths}.
As an independent test of the theory,
we also count statistics of
the linewidths of the excited regular modes.
 Conclusions and discussion follow in
Sec.~\ref{concl}.

\section{Theoretical model}
\label{Theory}
Classically, a
 deformed microcavity allows for both regular and chaotic motion (Fig.~\ref{SOS}),
 which are well separated, so that no trajectory can cross between them.
 In the quantum picture, however, it is possible for a wave living in one region
 to leak into the other via dynamical tunneling~\cite{davis81}, which introduces a
 coupling between regular and chaotic resonances.

\subsection{Mode-mode coupling theory}
\label{Anth}
The polarized field  [Transverse Electric (TE) or  Transverse Magnetic (TM)],
excited by the incident beam inside the microcavity, is written as a superposition of
one regular ($\omega$) and several ($n$) chaotic modes~\cite{AnYang}:
\begin{equation}
\psi(x,t)=\mathpzc{a}_{\omega
}(t)\mathpzc{c}_{\omega}(x)e^{ik_\omega z-i\omega t}+\sum_{n}\mathpzc{b}_{n}(t)
\mathpzc{c}_{n}(x)e^{ik_nz-i\omega_n t}.
\label{totfield}
\end{equation}
As said, the regular mode is coupled to the chaotic ones via dynamical tunneling, and therefore,
one can write a system of response equations~\cite{AnCoup} under the slowly-varying amplitude assumption~\cite{BoydNLO}, and integrate out the spatial part of the modes,
to obtain
\begin{subequations}
\begin{align}
\mathpzc{\ddot{b}}_n  + \omega_n^2\mathpzc{b}_n + \gamma_n\mathpzc{\dot{b}}_n
& = f_nE_0 - V_n\mathpzc{a}_\omega , \\
\mathpzc{\ddot{a}}_{\omega} + \gamma_\omega\mathpzc{\dot{a}}_\omega
+ \omega^2\mathpzc{a}_\omega & = \sum_nV_n\mathpzc{b}_n  .
\end{align}
\label{rspeqs}
\end{subequations}
Here $f_{n}$ is the coupling strength of the $n$-th chaotic mode with the
laser beam of amplitude $E_{0}$ and frequency $\omega_{0}$.
$V_n$ (assumed real) is the coupling strength of the
$n$-th chaotic mode with the regular mode, while
$\gamma_n$ and $\gamma_\omega$ are damping rates.
Assuming that $\omega_n\approx\omega_0$, one can  first set
\begin{subequations}
\begin{align}
b_n  &  = \mathrm{Re} \left[\mathpzc{b}_ne^{-i\omega_0t}\right],\\
a_{\omega} & =  \mathrm{Re} \left[ \mathpzc{a}_\omega e^{-i\omega_0t}\right],
\end{align}
\label{rwamp}
\end{subequations}
and then rewrite Eqs.~(\ref{rspeqs}) as~\cite{LiRC16}
\begin{subequations}
\begin{align}
\dot{b}_{n}+\gamma_{n}b_{n}  &  =f_{n}E_{0}-V_{n}a_{\omega},\\
\dot{a}_{\omega}+\left[  \gamma_{\omega}+i(\omega_{0}-\omega)\right]
a_{\omega}  &  ={\textstyle\sum\limits_{n}}V_{n}b_{n}.
\label{coup_osc}%
\end{align}
\end{subequations}
We are interested
in the steady-state solution,
obtained by setting $\dot{a}_{\omega}=\dot
{b}_{n}=0$.
The amplitude $a_\omega$ of the envelope of the regular mode
is found to be~\cite{LiRC16}
\begin{equation}
a_{\omega}=\frac{E_{0}\sum\limits_{n}f_{n}\frac{V_{n}}{\gamma_{n}}}{\left[
\gamma_{\omega}+i(\omega-\omega_{0})\right]  +\sum\limits_{n}\frac{V_{n}^{2}%
}{\gamma_{n}}} . \label{excit_amp}%
\end{equation}

\subsection{An alternative approach}
\label{fanoth}

The same equation may be derived in a different  way~\cite{Fano,QF_dyntun}.
Consider a
Hamiltonian $H_0$ modelling the closed billiard, whose eigenstates
$|a_\omega\rangle$ and $|b_n\rangle$
represent the regular- and chaotic states respectively, uncoupled to one another.
The coupling is introduced by the opening, that modifies the
Hamiltonian to the non-Hermitian $H=H_0+V$. Moreover, an incident beam
of amplitude $E$ is shone into the cavity.
We begin by writing an eigenfunction of  $H$ as the superposition
\begin{equation}
|\psi\rangle = a_\omega|a_\omega\rangle + \sum_n b_n|b_n\rangle + E|E\rangle.
\label{supstate}
\end{equation}
We have the following coupling properties:
\begin{eqnarray}
\nonumber
\langle a_\omega|H|a_\omega\rangle  &=& \omega - i\gamma_\omega  \\ \nonumber
\langle b_n|H|b_n\rangle &=& \omega_n - i\gamma_n \\ \nonumber
\langle a_\omega|H|b_n\rangle &=& V_n \\ \nonumber
\langle a_\omega|H|E\rangle &=& 0 \\
\langle E|H|b_n\rangle &=& f_n.
\label{couplings}
\end{eqnarray}
We also assume regular-, chaotic states, and state of the incident beam to
be orhtogonal to one another~\cite{biorthogonality}.
We can now take the Schr$\ddot{\text{o}}$dinger equation
\begin{equation}
H|\psi\rangle = \omega_0|\psi\rangle,
\label{SchrH}
\end{equation}
and sandwich it with
\begin{enumerate}
\item $\langle a_\omega|$, to obtain
\begin{equation}
\left(\omega-i\gamma_\omega\right)a_\omega
+ \sum_n b_nV_n = \omega_0 a_\omega
\label{sandwch1}
\end{equation}
\item$ \langle b_n|$, and we get
\begin{equation}
b_n = \frac{a_\omega V_n + f_nE}{\left(\omega_0-\omega_n\right)
+ i\gamma_n}.
\label{sandwch2}
\end{equation}
\end{enumerate}
We now plug Eq.~(\ref{sandwch2}) into Eq.~(\ref{sandwch1}) to obtain
\begin{equation}
a_\omega = \frac{E\sum_n\frac{f_nV_n}{(\omega_0-\omega_n) +
i\gamma_n}}{(\omega_0-\omega) + i\gamma_\omega
-\sum_n\frac{V^2_n}{(\omega_0-\omega_n) + i\gamma_n}}.
\label{aomega}
\end{equation}
As before, we assume that $\omega_n\simeq\omega_0$ for all chaotic states,
which simplifies Eq.~(\ref{aomega}) to the form
\begin{equation}
a_{\omega}=-\frac{E\sum\limits_{n}f_{n}\frac{V_{n}}{\gamma_{n}}}{\left[
\gamma_{\omega}+i(\omega_0-\omega)\right]  +\sum\limits_{n}\frac{V_{n}^{2}%
}{\gamma_{n}}}. \label{minexcit_amp}%
\end{equation}
Equation~(\ref{minexcit_amp}) differs from Eq.~(\ref{excit_amp}), previously derived, by
an overall minus sign, which however does not affect the excitation probability
(squared modulus of the amplitude), and by the $\omega_0-\omega$ term, where the two
frequencies are swapped with respect to Eq.~(\ref{excit_amp}).

 \subsection{Probability of excitation of a regular mode}
 \label{keysec}
 Let us now rewrite
\begin{subequations}
\begin{align}
\sum_{n}f_{n}\frac{V_{n}}{\gamma_n} & \simeq
n_{\gamma}\langle\frac{f_nV_n}{\gamma_n}\rangle, \\
\sum_{n}\frac
{V_{n}^{2}}{\gamma_{n}} & \simeq n_{\gamma}\langle\frac{V_n^{2}}{\gamma_n}\rangle,
\end{align}
\label{aversums}
\end{subequations}
where the averages are taken over $n_\gamma$ chaotic modes of small
enough linewidth ($\gamma$ sets the upper bound)
to effectively contribute to the excitation of the regular modes.
That way, we can express Eq.~(\ref{excit_amp}) as
\begin{equation}
a_\omega  = \frac{E_0n_{\gamma}\langle\frac{f_nV_n}{\gamma_n}\rangle}
{\left[\gamma_{\omega}+i(\omega-\omega_{0})\right]  +
 n_{\gamma}\langle\frac{V_n^{2}}{\gamma_n}\rangle}.
 \label{intm}
 \end{equation}
This can be rewritten (setting $\epsilon=\frac{E_0\langle f_nV_n/\gamma_n\rangle}
{\langle V_n^2/\gamma_n\rangle}$ and
$\Gamma~=~\frac{\gamma_\omega}{\langle V_n^{2}/\gamma_n\rangle}$) as
\begin{equation}
a_\omega  = \epsilon\frac{n_\gamma}
{\left[\Gamma +i\frac{(\omega-\omega_{0})}{\langle V_n^2/\gamma_n\rangle}\right]  + n_\gamma}.
\label{intm2}
\end{equation}
The excitation probability for the regular mode is therefore
\begin{equation}
|a_\omega|^2 = \epsilon^2\frac{n_\gamma^2}
{(\Gamma + n_\gamma)^2 + \frac{(\omega - \omega_0)^2}{\langle V_n^2/\gamma_n\rangle^2}},
\label{intm3}
\end{equation}
which becomes, at resonance, \cite{LiRC16}
\begin{equation}
|a_\omega|^2 = \epsilon^2\frac{n_\gamma^2}
{(\Gamma + n_\gamma)^2} .
\label{correct}
\end{equation}

\subsection{Statistics of chaotic states}
\label{statchastat}

Equation~(\ref{correct}) is central to our construction, as it links
the number of excited regular modes (proportional to $|a_{\omega}%
|^{2}$), that we measure directly, to the number of chaotic modes $n_{\gamma
}$, that we estimate as follows.

In principle, we lack information on the typical decay rate/linewidth of the
chaotic modes that contribute to the excitation of the regular ones.
Specifically, we do not know whether the latter decay within the Ehrenfest
time $\tau_{\mathrm{Ehr}}$ of quantum-to-classical correspondence, or whether they are, on average,
significantly longer lived (quasibound).
 Because of that, we present here three different models, the first entirely classical,
 the second based on the truncation of random unitary matrices~\cite{ZycSom}, suitable
 for quasibound states, and the third that combines the previous two~\cite{SchomTwor}, and
 that thus applies to an intermediate timescale.

 The following analysis refers to the classical dynamics of the
 chaotic billiard, and all the time-related quantities are
 expressed in units of the average (`Poincar\'e') time
 between two consecutive bounces of a ray on the boundary.
 \begin{enumerate}
\item  We begin with the classical description, and assume that the motion inside the chaotic
 part of the phase space is hyperbolic, so that the survival probability takes the form
 $P(t) \propto e^{-t/\tau_d}$, where $\tau_d$ is the mean dwell time
 of a trajectory in the system.  If there are $M$ states in the cavity at $t=0$,
 the average number of states that survive in the cavity by time 
$t^*<\tau_{\mathrm{Ehr}}$ is given by
\begin{equation}
n(t^*<\tau_{Ehr}) = 
Me^{-t^*/\tau_d}
= Me^{-1/\gamma\tau_d}  
\label{class_law}
\end{equation}
having set $\gamma=1/t^*$. 
In particular, the number of states that survive at the Ehrenfest time is given by
\begin{equation}
n(\tau_{Ehr}) = Me^{-t_{Ehr}/\tau_d} =
MN^{-1/\hat{\mu}\tau_d},
\label{Ehr_svvs}
\end{equation}
where $\hat{\mu}$ is of the order of the Lyapunov exponent, $N$
is the number of open channels, so that $\tau_d=M/N$, and we took
$\tau_{Ehr}=\hat{\mu}^{-1}\ln N$ (see Appendix~\ref{applyap} for details).
\item
The statistics of the spectrum of
a chaotic Hamiltonian is typically determined by means of
Random Matrix Theory (RMT)~\cite{stoeck}.
We now follow this approach in order to estimate
the number of long-lived states,
starting with an expression
for the probability distribution $P(r)$, $r=e^{-\gamma_n/2}$ ($\gamma_n$
escape rate of an eigenstate of the open system), obtained from truncated
random matrices~\cite{ZycSom}:
\begin{equation}
\Phi(r) = C\frac{2r}{\left(1-r^2\right)^2},
\label{Zycz_Somm}
\end{equation}
where $C$ is a normalization constant.
The number of eigenstates $n_{\gamma,\mathrm{RMT}}$   
with escape rate $\gamma_n<\gamma$
is then evaluated from the integral of Eq.~(\ref{Zycz_Somm})
\begin{equation}
n_{\gamma,\mathrm{RMT}} = \int_{r_\gamma}^{\sqrt{1/\tau_d}}
\Phi(r)dr 
\label{Pint}
\end{equation}
 with $r_\gamma =e^{-\gamma/2}$, under the assumptions that
$\tau_d\gg~1$, and
$\lim_{\gamma\rightarrow\infty}n_{\gamma,\mathrm{RMT}}  = M-N$,
that is the number of states that do not decay instantaneously.
The final result is
\begin{equation}
n_{\gamma,\mathrm{RMT}} \simeq M\left[1-\frac{1}{\tau_d}\frac{1}{1-e^{-\gamma}}\right].
\label{RMT_altog}
\end{equation}
\item
If we then want to  remove the states that decay within Ehrenfest time
from the estimate of $n_\gamma$,
 we just combine~(\ref{RMT_altog}) and~(\ref{Ehr_svvs}), obtaining~\cite{LiRC16,SchomTwor}
\begin{equation}
n_{\gamma,\mathrm{Weyl}} = \frac{M}{N^{1/\hat{\mu}\tau_d}}\left[1-\frac{1}{\tau_d(1-e^{-\gamma})}\right].
\label{ngam_law}
\end{equation}
The previous expression, which scales as a nonintegral power of the number of states
consistently with the fractal Weyl law~\cite{SchomTwor,NonZwor05},
is therefore a semiclassical correction to the RMT prediction. It depends on the Lyapunov exponent of the chaotic dynamics, and therefore it  takes into account system-specific properties.
\end{enumerate}
In what follows, we will validate Eq.~(\ref{class_law}), Eq.~(\ref{RMT_altog}), and Eq.~(\ref{ngam_law})
respectively against the experimental data.

\section{Chaotic ray dynamics and excitation of regular modes}
\label{chaotic_rays}
\begin{figure}[tbh!]
\includegraphics[width=8.5cm]{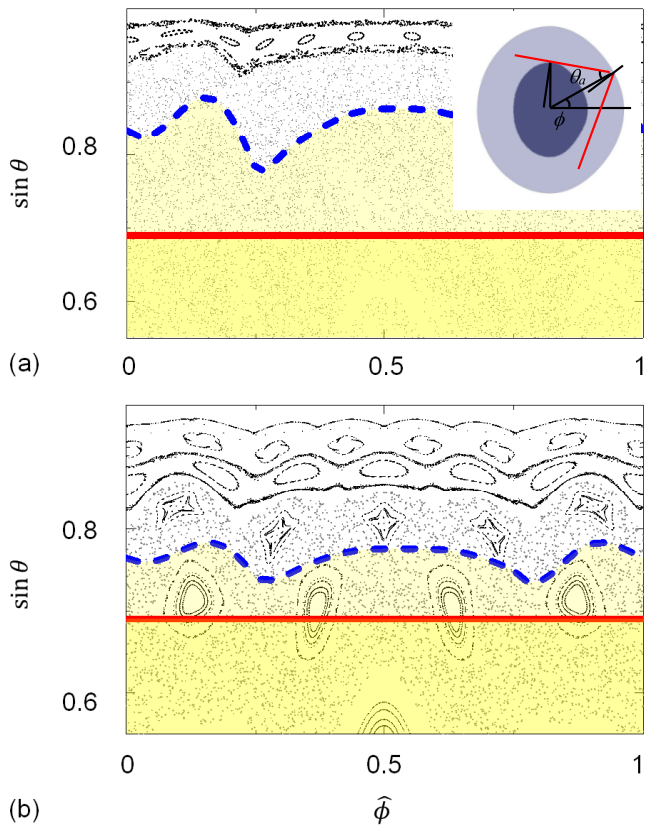}
\caption{(color online) (a) (inset) Sketch of the deformed microcavity with an inner absorber,
characterized by the angle $\theta_{a}$, and (main) corresponding
Poincar\'{e} surface of section
($\hat\phi\equiv\phi/2\pi$) with deformation factor
$\eta=11.7\%$.
The solid red line indicates the angle of total internal
reflection, while the dashed curve is given by an absorption angle $\theta_a$ such that
$r\simeq0.85$. Different shades of color indicate loss to the absorber
(lighter) and by refraction into air (darker). (b)
Poincar\'e section of the microcavity with $\eta=4.2\%$. The dashed curve is given by an absorption angle $\theta_a$ such that $r\simeq0.77$.}
\label{SOS}%
\end{figure}
\subsection{Absorber and phase space}
\label{absorber}
In order to achieve the full opening required to test the above predictions, we
introduce an absorber in the cavity.
In the analysis,
the dielectric microcavity (Fig.~\ref{SOS}(a), inset) has the deformed circle
$\rho(\phi)$ as boundary (see Sec.~\ref{expset} for details), which
encloses an absorber of shape $\rho(\phi)-R$. Figure~\ref{SOS} shows
the classical phase space, together with the critical line of total internal
reflection ($\sin\theta_{c}$), as well as the line given by the incidence
angle $\theta_{a}$, below which the reflected ray hits the absorber.
In what follows we neglect the dependence of $\theta_a$
on $\phi$ by taking the average value, approximately given
by the ratio $r$ of the mean radius of the absorber to the cavity's.
We assume in this model that the rays that hit the absorber are completely absorbed by it.
We will justify the assumption in Sec.~\ref{expset}.  
We also previously remarked that only a subset of  longer-lived chaotic states,
out of those available in the whole phase space, effectively contribute to the excitation
of the regular modes [Eqs.~(\ref{aversums})]. Because of that, we do not count
the rays that escape the cavity by refraction into the air with an angle of
incidence $\theta\ll\theta_{c}$, since these are
very lossy and they
are not expected to contribute to the excitation of the regular modes.
Instead, we only take into account the states supported on a strip of
the chaotic phase space with momentum above a certain threshold, $\sin
\theta>\sin\theta_{\mathrm{th}}$, to be chosen below but close enough to the
critical line of total internal reflection. Let us introduce the notation
$\xi\equiv\sin\theta_{a}-\sin\theta_{\mathrm{th}}$ to indicate the strip of
the phase space opened by the absorber.
The $N$ open channels (cf. Sec.~\ref{statchastat})
out of the $M$
Planck cells available in the phase space,
are produced by the absorber (full
opening, $N_{a}$) and the refraction out of the cavity (partial opening,
$N_{r}$), so that the mean dwell time of a ray is given by~\cite{LiRC16}
\begin{equation}
\tau_{d}=\frac{M}{N_{a}+N_{r}},
\label{T}%
\end{equation}
with $N_{r}=\frac{M}{A}\int_{\sin\theta_{a}}^{\sin\theta_{c}}T(\sin
\theta)d\sin\theta$, $T$ transmission coefficient according to Fresnel law, and
$A$ area of the chaotic phase space in exam, while $N_{a}=M\xi/A$.

The mean dwell time plays a central role in the present study, since the main idea of
our experiments resides in counting resonances as a function of the size of
the absorber, and therefore in using
$\tau_d$ as the variable for the
predictions~(\ref{class_law}),~(\ref{RMT_altog}), and~(\ref{ngam_law}).

\subsection{Excitation of the regular modes}
\label{WGMexct}
\begin{figure}[tbh!]
\includegraphics[width=8.5cm]{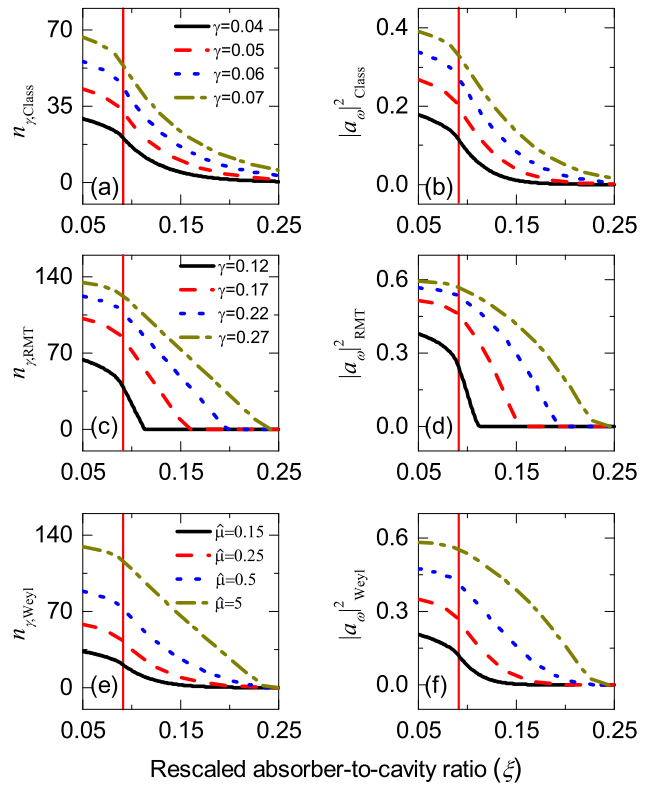}
\caption{(color online) (a), (c), (e) Number of chaotic states $n_{\gamma}$ vs. the
rescaled absorber-to-cavity ratio $\xi$, obtained by classical, RMT and Weyl law
respectively. (b), (d), (f) The corresponding expectations for $|a_{\omega}|^{2}$.
The red vertical line corresponds to the critical
angle, $\sin\theta_{c}\simeq0.69$. Here $\sin\theta_{\mathrm{th}}=~0.6$.
(c)-(f) adapted from Ref.~\cite{LiRC16}.}
\label{figuretwo}%
\end{figure}
We proceed by steps and examine the above theories
(classical, RMT, semiclassical) for the number of chaotic states $n_\gamma$
with escape rate less than $\gamma$, as a function of the mean dwell time
$\tau_d$, or, equivalently, the rescaled absorber-to-cavity ratio $\xi$.
We set
 $\hat{\xi}=1/\tau_d=\xi/A+N_{r}/M$,
 and rewrite the predictions of  section~\ref{statchastat} in terms of $\hat{\xi}$.
\begin{enumerate}
\item  The classical model [Eq.~(\ref{class_law})] becomes
\begin{equation}
n_{\gamma,\mathrm{Class}} = Me^{-\hat{\xi}/\gamma}
\label{nclass}
\end{equation}
in the new notation. Figure~\ref{figuretwo}(a) shows the decay
of the number of states as the opening increases in size, and [Fig.~\ref{figuretwo}(b)] the
correspondent decay of the probability of excitation of a regular mode:
while $n_{\gamma,\mathrm{Class}}$ and $|a_\omega|^2_{\mathrm{Class}}$
[obtained by plugging Eq.~(\ref{nclass}) into  Eq.~(\ref{correct})]
decrease slowly for $\xi$ small such that
$\theta_{a}<\theta_{c}$, that is when the loss is mainly due to refraction into air,
both quantities fall off rapidly, and nonlinearly, when the loss is entirely due to the
absorber (full opening).
 \item The RMT-based prediction, Eq.~(\ref{RMT_altog}), is also rewritten as
 a function of $\hat{\xi}$~\cite{LiRC16}:
\begin{equation}
n_{\gamma,\mathrm{RMT}}=M\left[  1-\frac{\hat{\xi}}{1-e^{-\gamma}}\right]  .
\label{RMT_x}%
\end{equation}
Its behavior 
is illustrated in Fig.~\ref{figuretwo}(c):
here $n_{\gamma,\mathrm{RMT}}$ decreases linearly with
$\xi$ in the region of total internal reflection, when the loss is entirely
due to the absorber. 
The probability of excitation
of the high-$Q$ regular modes $|a_{\omega}|^{2}_{\mathrm{RMT}}$ starts to fall
off as $\xi$ reaches some critical value, controlled by the parameter $\gamma$
[Fig.~\ref{figuretwo}(d)]. The other parameter
$\tilde{\Gamma}=\frac{\gamma_\omega}{M\langle V_n^{2}/\gamma_n\rangle}$
controls the slope of the curve.
\item
The semiclassical estimate~(\ref{ngam_law}) becomes, as a
function of~$\hat{\xi}$ \cite{LiRC16},
\begin{equation}
n_{\gamma,\mathrm{Weyl}}=\frac{M^{1-\hat{\xi}/\hat{\mu}}}{\hat{\xi}^{\hat{\xi}/\hat{\mu}}
}\left[  1-\frac{\hat{\xi}}{1-e^{-\gamma}}\right]  . \label{ngam_law_x}%
\end{equation}
The rescaled  Lyapunov exponent  $\hat{\mu}$ of the chaotic
region of the phase space
 is what really
characterizes~(\ref{ngam_law_x}), 
which resembles the linear RMT expression~(\ref{RMT_x})  for large enough $\hat{\mu}$, and otherwise
becomes visibly nonlinear [Figs.~\ref{figuretwo}(e)] when $\hat{\mu}\ll1$.
This nonlinearity produces a characteristic tail in the probability
$|a_\omega|^2_{\mathrm{Weyl}}$ [Fig.~\ref{figuretwo}(f)],
similar to that of the classical prediction~(\ref{class_law}). We therefore interpret it as a
signature of chaos,
which is
most evident slightly above the onset of chaotic dynamics.

\end{enumerate}
\subsection{Transient chaos}
\label{transch}
The survival probability  
leading to Eqs.~(\ref{class_law}) and~(\ref{ngam_law})
for the classical estimates
of the number of decaying states,
has an exponential
form because it rests on the assumption of
a fully chaotic phase space.
However, the phase portraits of Fig.~\ref{SOS} suggest the presence of
non-hyperbolic (`sticky') regions~\cite{ketzWeyl,IShudoSchom}, as well as of partial transport
barriers~\cite{MeissPBs,KetzPBs} even in the chaotic part of the phase space, which
would make the survival probability decay algebraically, instead of exponentially.
We address the issue by performing extensive ray-dynamics simulations of
the microcavity-shaped billiard of two different deformation factors, and computing
the survival probability in the chaotic region.  Here the absorber at the
center of the billiard constitutes the sole, full opening.
\begin{figure}[tbh!]
\includegraphics[width=8.5cm]{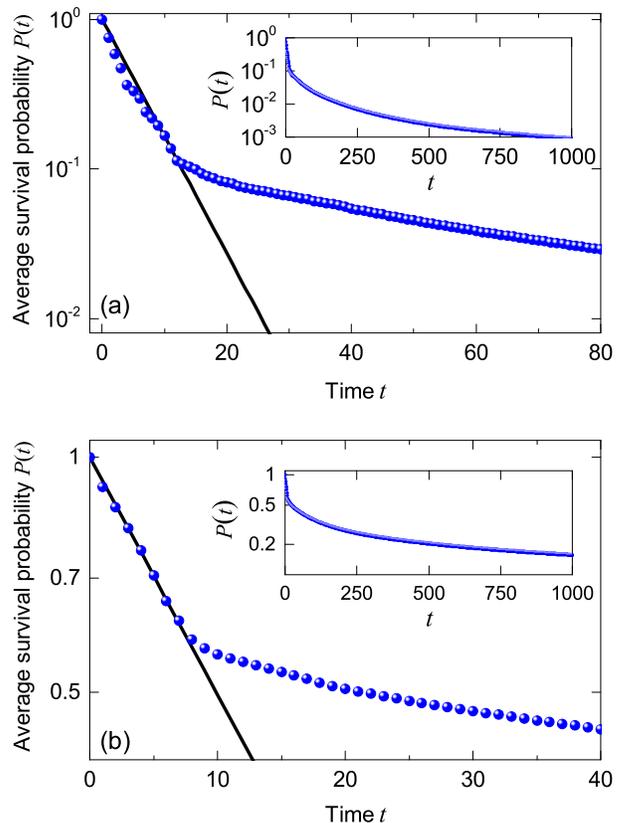}
\caption{(color online) Survival probability in the chaotic region (log scale). Points:
average survival probability $P(t)$ of a ray in the
microcavity 
vs. $t$ (in units of Poincar\'e time) at: (a) $\eta=11.7\%,\xi=0.13$, from $10^{6}$ randomly-started trajectories.
Line: $P(t)\propto\mathrm{exp}(-t/\tau_{d})$, $\tau_{d}=6$ (from Ref.~\cite{LiRC16});
(b) $\eta=4.2\%,\xi=0.1$, $\tau_d=14$.
 Insets: the long-time simulation showing algebraic decay.}%
\label{figurefive}%
\end{figure}
 Figure~\ref{figurefive} illustrates the results:
despite an
overall power-law decay, a closer look at the short-time dynamics reveals that
the decay is initially exponential,  behavior known as transient
chaos~\cite{LaiTel}.
The applicability of a model involving fully developed chaos
would depend on how the Ehrenfest time of quantum-to-classical correspondence
compares to the typical transition time $\tau_{\mathrm{trans}}$
by which the chaotic decay turns algebraic.
We estimated $\tau_{\mathrm{trans}}=6 (14)$ units of Poincar\'e time
 for a cavity with deformation
factor $\eta=11.7\% (4.2\%)$. We shall estimate the Ehrenfest times of the microcavity
of these deformations in Sec.~\ref{class_stat}, and confirm the validity of the fully
chaotic model for the present experiments.


\section{Experimental setup and measurement}
\label{expset}
\begin{figure}[tbh!]
\centerline{
\includegraphics[width=8.5cm,clip]{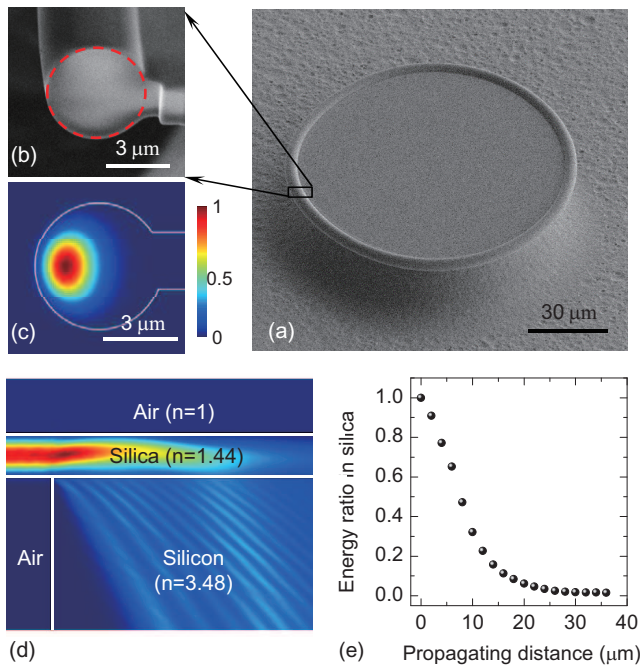}}
\caption{(a) Image of the microcavity obtained by Scanning Electron Microscopy (SEM).
(b) SEM cross-section image of the toroidal part, 
taken at an
angle of $56^{\circ}$ with the horizontal direction. (c)
Finite-element method simulation of a fundamental TE mode  (color scale in arbitrary units).
The white solid curve
is the boundary of the cavity. 
(d) Finite-element method simulation of the light propagating inside the $2$%
-$\mathrm{\mu}$\textrm{m}-thick silica waveguide bonding with a thick silicon
layer (color scale in arbitrary units). (e) Fraction of remaining energy in silica vs. the propagating distance.
}%
\label{cavity}%
\end{figure}

\begin{figure}[tbh!]
\includegraphics[width=8.5cm]{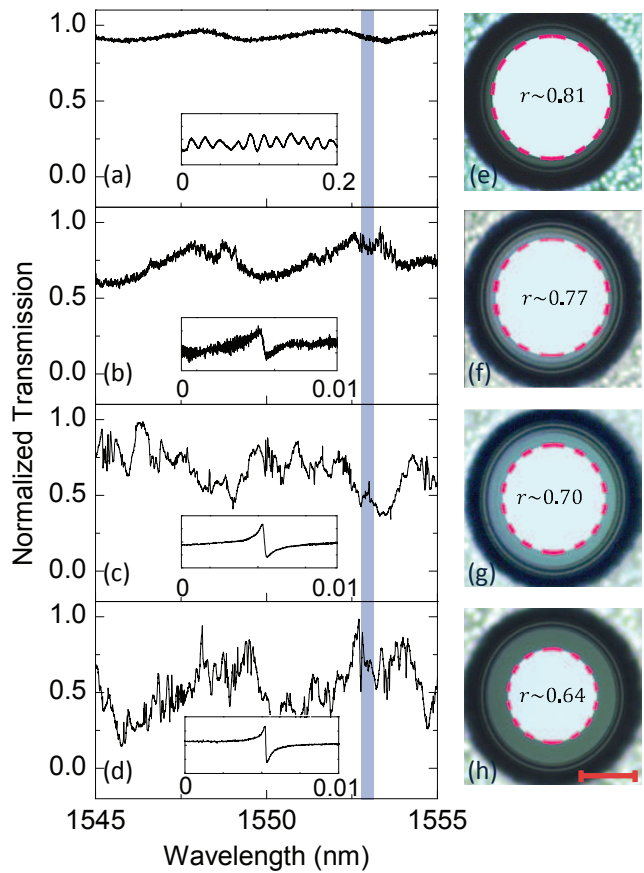}
\caption{(color online) Normalized
transmission and top-view optical images of the cavity with $r\simeq0.81$ [(a)
and (e)], $0.77$ [(b) and  (f)], $0.70$ [(c) and (g)], $0.64$ [(d) and (h)]. Inset of (a) shows background noise. Insets of (b)-(d) show the high-$Q$
modes. Reflection of the silica-to-silicon interface results in a brighter
color for the silicon pillar in the optical image (boundary shown by red
dashed curves). Scale bar is $50$ $\mathrm{\mu}$\textrm{m}.
Adapted from Ref.~\cite{LiRC16}.}%
\label{figurethree}%
\end{figure}

 The experimental apparatus consists of a deformed toroidal microcavity
of boundary shape given by the curve
\begin{equation}
\rho(\phi)=\left\{
\begin{array}
[c]{cc}%
\rho_{0}(1+\epsilon\sum_{i=2,3}a_{i}\cos^{i}\phi) & \mathrm{for}\cos\phi
\geq0,\\
\rho_{0}(1+\epsilon\sum_{i=2,3}b_{i}\cos^{i}\phi) & \mathrm{for}\cos\phi<0,
\end{array}
\right. \label{df_circ}%
\end{equation}
with $\rho_0= 60$ $\mathrm{\mu}$\textrm{m},
$a_{2}=-0.1329,a_{3}=0.0948,b_{2}=-0.0642$, and $b_{3}=-0.0224$. The WGMs in
the deformed microcavity have been demonstrated to possess ultrahigh quality
factors in excess of $10^{8}$ in the $1550$ nm wavelength band and to exhibit
highly directional emission towards the $180^{\circ}$ far-field direction,
which emits tangentially along the cavity boundaries at polar angles $\phi
=\pi/2$ and $\phi=3\pi/2$~\cite{xiao13}. The deformation is controlled by $\eta=(d_{\max
}-d_{\min})/d_{\max}$, $d_{\max}$ and $d_{\min}$ respectively the maximum and
minimum diameters of the cavity. The parameter $\eta$ is related to $\epsilon$
through $\eta=\epsilon\left\vert a_{2}+a_{3}+b_{2}-b_{3}\right\vert /2$.
The microcavity is
coupled to a free-space propagating laser beam of wavelength $\lambda\simeq1550$ \textrm{nm}
(swept from 1555 \textrm{nm} to 1545 \textrm{nm}, free-spectral range 4.4 \textrm{nm} \cite{FSEdef}),
or $635$ \textrm{nm}
(swept from 639 \textrm{nm} to 637 \textrm{nm}, free-spectral range 0.74 \textrm{nm}),
as shown in Fig.~\ref{figureone}.
The microtoroid [refractive index $\simeq1.44,1.46$ depending on
$\lambda$, Fig.~\ref{cavity}(a)] has principal/minor diameters of $120/5$ $\mathrm{\mu}$\textrm{m},
consistently with the two-dimensional model.
Thus the effective
Planck constant $h_{\mathrm{eff}}\sim\lambda/a\sim10^{-2}$ ($a$: principal
diameter) justifies the semiclassical analysis.  
The microcavity is fabricated through optical lithography, buffered \ce{HF} wet etching,
\ce{XeF2} gas etching, and \ce{CO2} pulse laser irradiation. The resulting
silica microtoroid is supported by a silicon pillar of similar shape, which has a high refractive index ($\simeq3.48,3.88$),
and it acts as the absorber in the model.
After each measurement of the transmission spectrum (Fig.~\ref{figurethree}),
the top diameter of the silicon pillar, connected with the silica disk,
is progressively reduced by a new isotropic \ce{XeF2} dry etching process.
In this way we control the openness of the
microcavity with the ratio $r$ between the top diameters of pillar and toroid.
 Finite-element method simulations [Fig.~\ref{cavity}(d),(e)] show that the light power
decreases to less than $5\%$ of the input value, when propagating by a
distance of $20$ $\mathrm{\mu}$\textrm{m} inside the $2$-$\mathrm{\mu}%
$\textrm{m}-thick silica waveguide bonding with a silicon wafer,  
as it is reasonable to expect, given the high refractive index of the silicon.
Thus the silicon pillar acts as a full absorber,
consistently with the model presented here.
On the other hand, high-$Q$ regular modes
living inside the toroidal part, whose circular cross section has diameter of $5$
$\mathrm{\mu}$\textrm{m}, do not leak into the silicon pillar and therefore
are not directly affected by the pillar size. Figure~\ref{cavity}(c) illustrates
the numerical simulation of a  regular TE mode, that is confined in the toroidal
region.
Due to the free-space propagation, the laser beam can only enter the cavity
with a relatively large angle of incidence,  which results in smaller angles of refraction
into the resonator, and of incidence with its boundary at the next collisions.
As a consequence, the laser beam only directly excites the
chaotic cavity modes localized in the central region of the cavity, which
in turn couple with the regular modes localized in the outer toroid via
dynamical tunneling, consistently with the model of Sec.~\ref{Theory}.

The dependence of the
transmission spectra on the pillar size is shown in
Fig.~\ref{figurethree}. When the pillar approaches the inner edge of the
toroid (Figs.~\ref{figurethree}(a) and \ref{figurethree}(e), $r\simeq0.81$), no high-$Q$
regular modes are observed in the spectrum, since most of the
probe laser field in the cavity radiates into the silicon and cannot tunnel to couple with
high-$Q$ regular modes. As we gradually reduce the size of the pillar
[Fig.~\ref{figurethree}(f), $r\simeq0.77$], increasingly many high-$Q$ modes
appear in the spectrum [Fig.~\ref{figurethree}(b)]. When the
absorber-to-cavity ratio $r$ is small enough (Figs.~\ref{figurethree}(g) and
\ref{figurethree}(h), $r\lesssim0.7$), the transmission no longer changes sensibly
[Figs.~\ref{figurethree}(c) and \ref{figurethree}(d)], and the number of high-$Q$ modes in the
spectrum also stabilizes.

It is noted that the high-Q regular modes are easily recognized even when the coupling efficiency is low, because linewidths coming from noise are typically orders of magnitude wider than those of the high-Q resonances in the transmission spectrum, as shown in the insets of [Fig.~\ref{figurethree}(a) and (b)]. That is to say, nearly all the existing high-Q modes are conspicuous in the spectra and can be detected.

\section{Statistics of Chaotic Resonances}
\label{results}
As anticipated, we use the transmission spectra to test the theory, by
counting the excited high-$Q$ regular modes for different sizes of the silicon
pillar. 
 A polarization controller is used to alternatively excite TE- or TM modes, which
are collected by the photon receiver, and read from the transmission spectra.
We single out and add up the modes with high $Q$ factors ($Q>10^5$)  for
both polarizations, 
and then multiply the
result by the ratio of free-spectral range to the range of wavelengths swept by the laser beam
(for example, that is $0.74/(639-637)$ for visible light). 
Since
TE-  and TM modes are not perfectly orthogonal to each other in the real microcavity, some may be counted twice, which is the
main source of uncertainty in our data.

Based on the assumption (discussed in Sec.~\ref{linewidths}) that the number of
regular modes excited via dynamical tunneling is proportional to the probability
of excitation of a single regular mode, given by  Eq.~(\ref{correct}), we  henceforth
test all the predictions presented in the theoretical sections, and plug Eqs.~(\ref{nclass}),
Eq.~(\ref{RMT_altog}), and Eq.~(\ref{ngam_law}) respectively into the expectation
\begin{equation}
n_{\mathrm{reg}} =   \kappa\frac{n_\gamma^2}
{\left(\Gamma+ n_\gamma\right)^2}
\label{nregmodes}
\end{equation}
for the counted high$-Q$ resonances.

\subsection{Classical model}
\label{class_stat}
\begin{figure}[tbh!]
\centerline{
\includegraphics[width=9cm]{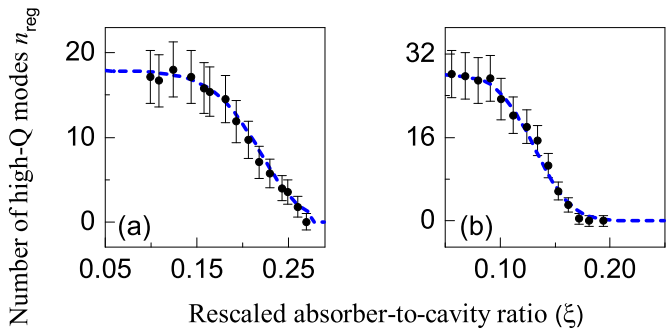}}
\caption{(color online) Dots: number of high-$Q$ regular modes ($n_{\omega}$)
observed in the transmission spectra of the microcavity coupled to infrared light,
as a function of rescaled absorber-to-cavity ratio $\xi$. Blue dashed curve: best fit of the
classical prediction~(\ref{nclass})[together with Eq.~(\ref{nregmodes})].
Left: $\eta=11.7\%$; right: $\eta=6\%$.
Here $\sin\theta_{c}\simeq0.69$
and $\sin\theta_{\mathrm{th}}=0.6$.}
\label{figureclass}%
\end{figure}

We
start with the classical
model. Equation~(\ref{nclass}) is plugged into Eq.~(\ref{nregmodes}), and fitted to the data
via the parameters $\tilde{\Gamma}$, $\gamma$, up to an overall multiplicative
constant.  The total number of
chaotic states is estimated theoretically as $M\simeq A/h_{\mathrm{eff}}$ ($A$
area of the chaotic phase space we consider).
The results are shown in Fig.~\ref{figureclass}, with details in Table~\ref{table:class}.
\begin{table} [tbh!]
\centering
\caption[classfit]{
Parameters  related to the best-fit of  Eqs.~(\ref{nclass}) and~(\ref{nregmodes}) to the data,
and to the experimental conditions. Here
$\gamma$ is expressed in units of $T^{-1}$,  with $T\simeq3\cdot10^{-13}$s Poincar\'e time.}
\begin{tabular} {c c c c c c}
\hline\hline
$\tilde{\Gamma}$ &  $\gamma$ &  $\eta$ & $\lambda$(nm) & $M$  & $\chi^2$ \\ [0.5ex]
\hline
6$\cdot10^{-5}$  &          0.01     &   4.2\%   & 630   & 40   &  4	   \\
2$\cdot10^{-4}$ & 	      0.015     &   4.2\%  &  1550  &  		   20   &    1.6      \\
6$\cdot10^{-4}$ & 	      0.014     &   6.0\%  &   630  &  		   40 	  &  3.1   \\
$10^{-3}$ & 	      0.017     &   6.0\%  &   1550  &  		   20    &  1.1  \\
2.4$\cdot10^{-4}$ &         0.017     &  11.7\%  &   1550  &          50         &      0.4 \\
\hline
\end{tabular}
\label{table:class}
\end{table}
The classical prediction appears to fit the data rather well, overall.
The statistical test of $\chi^2$~\cite{chi-s} evaluates
the average discrepancy between expectations [$n_{\gamma,\text{Class}}(\xi_i)$]
and observations ($n_{\gamma,i}$), divided by
the experimental errors $\sigma_i$, in $d$ degrees of freedom:
\begin{equation}
\chi^{2}=\frac{1}{d}\sum_i\frac{\left[n_{\gamma,i}-n_{\gamma,\text{Class}}(\xi_i)%
\right]^{2}}{\sigma_i^{2}}.
\end{equation}
Here we generally obtain
$\chi^2\approx1$, indicating  that the extent of the match
between observations and estimates is in accord with the error variance.
 However, in order for
 this model to be accurate, all the chaotic states indirectly detected by the experiment should
 decay within Ehrenfest time, which seems unlikely, in principle.
 The minimum decay time of the chaotic modes is determined
 from the best fits as $\gamma^{-1}$, and it can be compared with
 the Ehrenfest time (we should find $\tau_{\mathrm{Ehr}}>\gamma^{-1}$),
 with the latter estimated in terms of the laser wavelength, the
 dimensions of the cavity, the Lyapunov exponent of the classical
 dynamics, and the size of the absorber (see Appendix~\ref{app:Ehr}
 for details).
In this regard, the absorber-to-cavity ratio $\xi$ varies within a certain range,
thus we estimate
upper and lower bounds for $\tau_{\mathrm{Ehr}}$  in units of the Poincar\'e time $T$, as summarized
in Table~\ref{table:ET}, for cavities of different deformations, coupled to either
visible or infrared light.
 \begin{table} [tbh!]
\centering
\caption[ET]{
Lower and upper bounds for the typical Ehrenfest time (in units of
Poincar\'e time) of the chaotic states
of microcavities with different deformation factors, coupled to either infrared or
visible light, in comparison with the average $\gamma^{-1}_{\mathrm{Class}}$ coming from Table~\ref{table:class}, and $\gamma^{-1}_{\mathrm{Weyl}}$
coming from Table~\ref{table:Weyl}.}
\begin{tabular} {c c c c c c}
\hline\hline
 $\eta$ & $\lambda$(nm) & $\tau_{Ehr}^{min}$ & $\tau_{Ehr}^{max}$
 & $\gamma_{\mathrm{Class}}^{-1}$ & $\gamma_{\mathrm{Weyl}}^{-1}$ \\ [0.5ex]
\hline
4.2, 6\%   &        630        &    6.5	   &       14   &		83        &  7.4\\
 4.2, 6\%  &       1550       &  	 15 	    &	    25	      &		63        &    6.3 \\
11.7\%  &          1550       &          10       &        14      &           59     &     5.3\\
\hline
\end{tabular}
\label{table:ET}
\end{table}
As we can see, the estimated Ehrenfest time of the rays in the microcavity is always
significantly shorter than the minimal escape time $\gamma^{-1}$, although of the same order of
magnitude. That suggests that the classical model for the statistics of the chaotic states
alone does not describe the experiment consistently with the assumptions.

\subsection{RMT- and semiclassical predictions}
\label{RMTSC}
\begin{widetext}

\begin{figure}[tbh!]
\centerline{
\includegraphics[width=.95\textwidth]{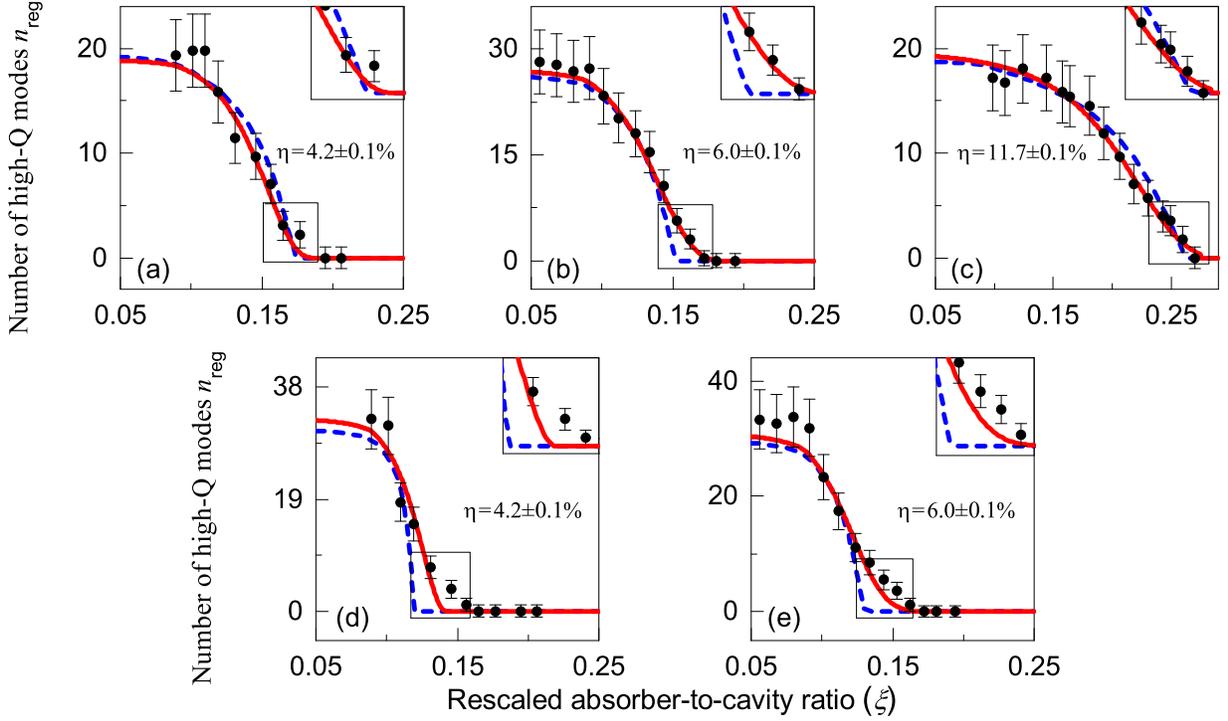}}
\caption{(color online) Number of high-$Q$ regular modes ($n_{\mathrm{reg}}$)
observed in the transmission spectra of the microcavity (dots), as a function
of rescaled absorber-to-cavity ratio $\xi$. Blue dashed- and red solid curves
are respectively RMT- and semiclassical prediction best fits.
(a), (b), (c): infrared light; (d), (e): visible light.
Here $\sin\theta_{c}\simeq0.69$
and $\sin\theta_{\mathrm{th}}=0.6$. Insets: the area where the two curves
differ most.}%
\label{moreexperimentdata}%
\end{figure}

\end{widetext}

Next, we validate $i)$ the
RMT-based prediction~(\ref{RMT_x}), and $ii)$ the semiclassical
expression~(\ref{ngam_law_x}), which we alternatively plug into
Eq.~(\ref{nregmodes}).
 The results are illustrated in Fig.~\ref{moreexperimentdata} (details in Tables~\ref{table:RMT}
and~\ref{table:Weyl}),
 for two microcavities of
distinct deformations, probed at visible and infrared wavelengths.

In the RMT-based approach
we have two
fitting parameters, $\gamma$ and $\tilde{\Gamma}$.
\begin{table} [tbh!]
\centering
\caption[RMTfit]{
 Parameters  related to the best-fit of  Eqs.~(\ref{RMT_x}) and~(\ref{nregmodes}) to the data,
and to the experimental conditions in Fig.~\ref{moreexperimentdata}.
Here $\gamma$ is expressed in units of $T^{-1}$,  with $T\simeq3\cdot10^{-13}$s Poincar\'e time.}
\begin{tabular} {c c c c c c}
\hline\hline
$\Gamma$ &  $\gamma$ &  $\eta$ & $\lambda$(nm) & $M$ & $\chi^2$   \\ [0.5ex]
\hline
0.025  &          0.1     &   4.2\%   & 630   &    40  &   16	   \\
0.08 & 	      0.15     &   4.2\%  &  1550  &  		   20   & 3.2   \\
0.07 & 	      0.11     &   6.0\%  &   630  &  		   40 	  & 11   \\
0.08 & 	      0.13     &   6.0\%  &   1550  &  		   20 &  2     \\
0.11 &         0.17     &  11.7\%  &   1550  &          50  &  1.1      \\
\hline
\end{tabular}
\label{table:RMT}
\end{table}
Figure~\ref{moreexperimentdata} shows overall
agreement between the experimental data  and this theory,
particularly in the infrared band and
at large
deformation.
However, the purely RMT model
 fails to capture the tail of the data at larger sizes of the absorber
in the experiments with visible light,
where the
$\chi^2$ 
significantly exceeds the optimal
value of unity.

\begin{table} [tbh!]
\centering
\caption[Weylfit]{
 Parameters  related to the best-fit of  Eqs.~(\ref{ngam_law_x}) and~(\ref{nregmodes}) to the data,
and to the experimental conditions in Fig.~\ref{moreexperimentdata}.
Here $\gamma$ and $\mu$ are expressed in units of $T^{-1}$,  with $T\simeq3\cdot10^{-13}$s Poincar\'e time.}
\begin{tabular} {c c c c c c c}
\hline\hline
$\Gamma$ &  $\gamma$ &  $\eta$ & $\lambda$(nm) & $M$ & $\mu$ & $\chi^2$   \\ [0.5ex]
\hline
0.24  &          0.12     &   4.2\%   & 630   &    40  &   0.05 & 6.9	   \\
0.28 & 	      0.16     &   4.2\%  &  1550  &  20   & 0.05 &  1.5   \\
0.55 & 	      0.15     &   6.0\%  &   630  &   40 	  & 0.05 & 4.8   \\
0.54 & 	      0.16     &   6.0\%  &   1550  &  20 & 0.05  & 0.6      \\
0.73 &         0.19     &  11.7\%  &   1550  &    50  &  0.1 &  0.5      \\
\hline
\end{tabular}
\label{table:Weyl}
\end{table}
\begin{figure}[tbh!]
\centerline{
\includegraphics[width=.5\textwidth]{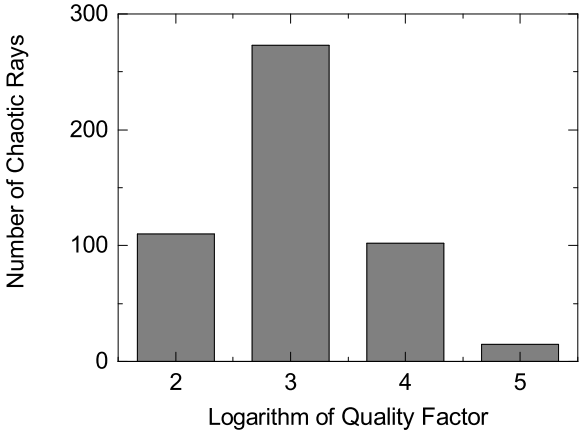}}\caption{Statistics of
the quality factor for the chaotic rays in the deformed microcavity, from a
ray-dynamics simulation.}%
\label{Qstat}%
\end{figure}
After that,
the semiclassical correction~(\ref{ngam_law_x}) is tested,
using the finite time Lyapunov exponent $\hat{\mu}$ evaluated by direct iteration, over a short enough time for the
dynamics to be still hyperbolic (cf. Sec.~\ref{transch}).
In addition, we still have
the estimated parameter $M$ and the fitted parameters $\gamma$ and $\Gamma$.
This expression is found in better agreement with the experimental data ($\chi^{2}\approx1$)
than the RMT-based estimate at smaller deformation and in the visible light band,
where the two predictions differ the most due to the smaller $\hat{\mu}$ [cf. Fig. \ref{figuretwo}].
Specifically, the semiclassical theory
accounts for the tail of the curve, that corresponds to the microcavity
having the largest openings and thus with the maximum number of instantaneous
decay states, where the
semiclassical correction is important.

Similarly to  the classical model, we check whether the fitted values of the parameter $\gamma$
for the RMT and semiclassical expressions
make physical sense. It is found that, typically,
$\gamma\simeq0.15$; now recall that
$\gamma^{-1}=\tau_{esc}$
the minimum escape time
of the chaotic rays contributing to the excitation of the regular modes,
from which $Q=2\pi\nu\tau_{esc}\sim10^{3}$, on average ($\nu$ is the frequency of the laser beam).
We find this estimate consistent with the typical order of $Q$ independently
obtained from ray-dynamics simulations (Fig.~\ref{Qstat}),
which corroborates the result from the analysis.
By the same token, one can write the linewidth of a resonance as
Im $\Omega = -\frac{a}{2c}\gamma$, where again $a$ is the cavity radius
and $c$ the speed of light inside the silica. We estimate on average
Im~$\Omega\simeq-0.1$, which is close to the median value
Im~$\hat{\Omega}\simeq-0.15$ of the distribution of resonances
 found in the numerical experiment
of ref.~\cite{WierMain}, where a stadium-shaped microcavity
of refractive index $n=1.5$ was considered.
At last, but importantly, we compare $\gamma^{-1}$ with the
Ehrenfest time. Equations~(\ref{RMT_x}) and ~(\ref{ngam_law_x}) are based on the
assumption $\tau_{Ehr}<\gamma^{-1}$, the opposite of the classical model's.
We find from our fits (Tables~\ref{table:RMT} and~\ref{table:Weyl}) that
$\tau_{Ehr}\approx\gamma^{-1}$ in all the realizations of the experiments,
and therefore the above assumption is not always validated within the uncertainties.
We believe at the present stage
the semiclassical prediction to be a more accurate model for the statistics of the chaotic states
than the entirely classical one.
All the same,
a number of chaotic states that escape whithin Ehrenfest time may also contribute to the excitation
of the regular modes,
for our experiment to capture
that intermediate time scale at the border line between ray- and purely wave-like modes.
One could at that point
combine theories using the following expression~\cite{SukJacq}:
\begin{equation}
n_\gamma = \varepsilon n_{\gamma,\mathrm{Class}} + (1-\varepsilon)n_{\gamma,\mathrm{Weyl}},
\label{comb}
\end{equation}
which would, however, add one presently unknown parameter ($\varepsilon$) to the analysis.
\section{Tunneling rates and statistics of linewidths}
\label{linewidths}
In this section, we discuss the proportionality between the
probability of excitation of a single regular mode and the number
of excited regular modes in the microcavity, which we have
estimated as
\begin{equation}
n_{\mathrm{reg}} =   \kappa\frac{n_\gamma^2}
{\left(\Gamma+ n_\gamma\right)^2}.
\tag{\ref{nregmodes}}
\end{equation}
In doing so, we have implicitly neglected the $\omega$~-~dependence
of the excitation probability
of  a regular mode
at resonance
\begin{equation}
\nonumber
|a_\omega|^2 = \epsilon^2\frac{n_\gamma^2}
{(\Gamma + n_\gamma)^2} .
\tag{\ref{correct}}
\end{equation}
In order to better understand this approximation, let us
restart from the excitation amplitude
\begin{equation}
\nonumber
a_{\omega}=\frac{E_{0}\sum\limits_{n}f_{n}\frac{V_{n}}{\gamma_{n}}}{\left[
\gamma_{\omega}  +i(\omega-\omega_{0})\right]\sum\limits_{n}\frac{V_{n}^{2}%
}{\gamma_{n}}} .
\tag{\ref{excit_amp}}%
\end{equation}
Here, we can regard the term
\begin{equation}
\gamma_\omega^{\mathrm{tot}} = \gamma_{\omega} +
\sum_n\frac{V_n^2}{\gamma_n}
\label{step_back}
\end{equation}
 as the total (hence measured) linewidth~\cite{AnYang},
where the first term $\gamma_\omega$ indicates the intrinsic linewidth of the regular mode, while the
second,  $\gamma_\omega^{\mathrm{dyn}}\equiv\sum_n\frac{V_n^2}{\gamma_n}$,
represents the decay rate into the chaotic modes.
It is to be determined whether
the fluctuations ultimately due to the dependence of these quantities on
the frequency $\omega$ of the mode can be neglected, say, to a first-order
approximation.

In what follows, we first argue that such fluctuations are small compared to the
decay rates and therefore the linewidths of the regular modes,
based on a semiclassical treatment of dynamical tunneling, and secondly
we test the validity of the approximation~(\ref{nregmodes}) with
two independent experiments.

\subsection{Action-based prediction of tunneling rates}

The tunneling rates into the chaotic field of distinct regular WGMs localized in the
toroid of the microresonator, or equivalently, in the top region of the phase portraits
of Fig.~\ref{SOS}, vary with the momentum of the corresponding rays, as it can be inferred from the
expression for the penetration through a potential barrier~\cite{Landau}
\begin{equation}
\gamma^{\mathrm{tun}} \propto e^{-\frac{2}{\hbar}\int_a^b |p|dq} .
\label{bartun}
\end{equation}
Intuitively, since the WGMs are confined in a narrow strip of the phase space
(cf. Fig.~\ref{SOS}),
the distributions of momenta of the regular trajectories
and of their tunneling rates into the chaotic sea are also supposedly quite narrow.
For an estimate of the variation of the tunneling rate of a regular mode
with the momentum, we need to be more accurate, and we may use
an expression derived in~\cite{Backer1} for the tunneling rate
out of a stability island into the chaotic region of a mixed phase space
\begin{equation}
\gamma_\omega^{\mathrm{dyn}} = \frac{c}{\sqrt{1-S_\omega}}
e^{-\frac{2A_{\mathrm{reg}}}{h_{\mathrm{eff}}}
\left\{\sqrt{1-S_\omega}-S_\omega\ln\left(\frac{1+\sqrt{1-S_\omega}}{\sqrt{S_\omega}}\right)\right\}},
\label{BackRate}
\end{equation}
where 
$c$ is a constant, while
$S_\omega = {A_{\mathrm{reg}}^{-1}}\oint pdq$
is the quantized action of the classical orbit corresponding to the regular mode in
exam, scaled by   the area of the regular region of the phase space $A_{\mathrm{reg}}$.
In the present construction $S_\omega<1$, which is also true in our setting.
Then, by Taylor-expanding  the action, the overall expression for the tunneling
rate becomes
\begin{equation}
\gamma_\omega^{\mathrm{dyn}} \approx
ce^{-\frac{2A_{\mathrm{reg}}}{h_{\mathrm{eff}}}
\left\{1+\frac{S_\omega}{2}\ln S_\omega\right\}} .
 \label{AppRate}
 \end{equation}
 At this point, one can compute with some algebra the differential
 \begin{equation}
d\gamma_\omega^{\mathrm{dyn}} \approx -\gamma_\omega^{\mathrm{dyn}}\frac{A_{\mathrm{reg}}}{h_{\mathrm{eff}}}\ln S_\omega dS_\omega,
\label{DiffAppRate}
\end{equation}
and thus an estimate for the relative error of the tunneling rate with
the change in action
\begin{equation}
\frac{\Delta\gamma_\omega^{\mathrm{dyn}}}{\gamma_\omega^{\mathrm{dyn}}} \approx
-\frac{A_{\mathrm{reg}}}{h_{\mathrm{eff}}}\ln S_\omega \Delta S_\omega .
\label{findiffapprate}
\end{equation}
Recalling the definition of action, and, in particular, that the WGMs in the
microcavity are supported on regular orbits in the upper part of the phase space
that closely follow the boundary,
where the momentum $p$ is almost constant along each trajectory,
we may write the change in action as
\begin{equation}
\Delta S_\omega \simeq \frac{2\pi\Delta p}{A_{reg}},
\label{chact}
\end{equation}
whence the estimate  $\frac{\Delta\gamma_\omega^{\mathrm{dyn}}}{\gamma_\omega^{\mathrm{dyn}}} \sim 10^{-1}$ in our experimental
conditions.

\subsection{Statistics of the regular modes and their linewidths}

The expression
\begin{equation}
\gamma_\omega^{\mathrm{tot}} = \gamma_{\omega} +
\sum_n\frac{V_n^2}{\gamma_n}
\simeq
\gamma_{\mathrm{reg}} + n_\gamma\left\langle\frac{V_n^2}{\gamma_n}\right\rangle,
\label{twostepsback}
\end{equation}
 indicates
that $\gamma^{\mathrm{tot}}_{\omega}$ increases with $n_\gamma$.
We shall now neglect the dependence 
on  $\omega$,
and derive an expression
for $\gamma^{\mathrm{tot}}_{\mathrm{reg}}$ in terms of
the number of regular modes $n_{\mathrm{reg}}$,
to be tested with an experiment.
Let us first
write $n_\gamma$ as a function of the observed quantity $n_{\mathrm{reg}}$,
by solving the quadratic equation~(\ref{nregmodes})
\begin{equation}
n_\gamma = \Gamma\frac{n_{\mathrm{reg}} + \sqrt{\kappa n_{\mathrm{reg}}}}
{\kappa-n_{\mathrm{reg}}}.
\label{nomngam2}
\end{equation}
Recalling the definition $\Gamma=~\frac{\gamma_\omega}{\langle V_n^{2}/\gamma_n\rangle}$, we have
\begin{equation}
\gamma_{\mathrm{reg}}^{tot} = \gamma_{\mathrm{reg}}\left[1+
\frac{n_{\mathrm{reg}} + \sqrt{\kappa n_{\mathrm{reg}}}}
{\kappa-n_{\mathrm{reg}}}\right].
\label{gamtot}
\end{equation}
\begin{figure}[tbh!]
\centerline{
\scalebox{1.2}{\includegraphics{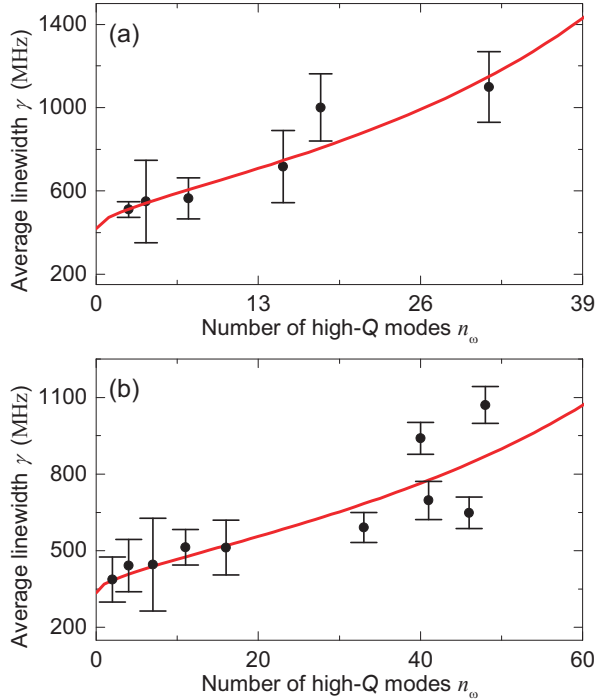}}}
\caption{Dots: average linewidth of the excited regular modes vs. their number
from a free-space coupling experiment (cf. Fig.~\ref{figurethree}).
Each data
point represents one experiment with a different size of the silicon pillar.
Solid line: best-fit curve of Eq.~(\ref{gamtot}). (a)
 $\eta=4.2\%$, $\lambda=~635$~nm,
with fitting parameters $\gamma_{\mathrm{reg}} = 419$MHz, and $\kappa = 78$.
 (b)
$\eta=6\%$, $\lambda=~635$~nm,
$\gamma_{\mathrm{reg}} = 378$MHz, and $\kappa = 120.5$.}
\label{gamvsn}
\end{figure}
We may now fit this prediction to the data in the free-space coupling
experiment, that is the
number of detected high-$Q$ modes, and their linewidths.
Figure~\ref{gamvsn} shows that 
the data points representing the
 average linewidths of the regular modes
 do follow
the trend predicted by 
Eq.~(\ref{gamtot}). 
 
\begin{figure}[tbh!]
\centerline{
\includegraphics[width=.5\textwidth]{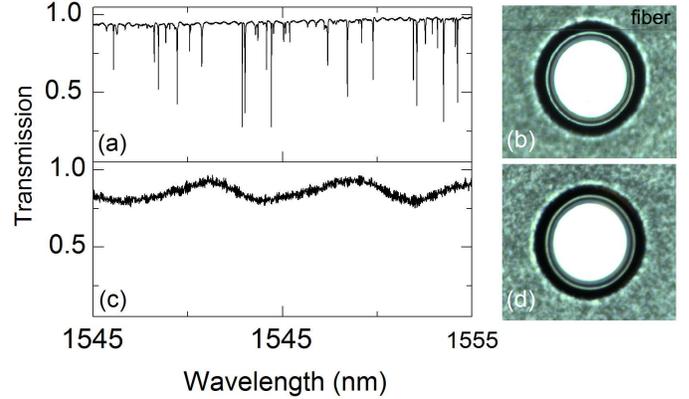}}\caption{ Normalized
transmission and top-view images of the cavity coupled by fiber taper [(a) and
(b), notice the fiber beside the cavity] and free-space laser beam [(c) and (d)].
The resonances in (a) have
$Q$ factors typically of the order of $10^6-10^7$.
Here the absorber-to-cavity ratio is
$r\simeq0.83$.}%
\label{fiber_taper}%
\end{figure}

 Although the
 fluctuations can be significant here [Fig.~\ref{gamvsn}(b)], 
overall quantitative consistency between the proposed model
and the data is definitively found through an auxiliary experiment,
illustrated in
Fig.~\ref{fiber_taper}(a)-(b), and
described as follows.  
 The WGMs are excited directly (no excitation of chaotic modes)
through a tapered fiber~\cite{VahFib00}. The silicon pillar attached to the
microtoroid has largest size, so that dynamical tunneling is inhibited and no
whispering-gallery mode can be excited with the free-space coupling
[Fig.~\ref{fiber_taper}(c)].
Measuring linewidths of the detected modes
results in a quality factor $Q_{\mathrm{reg}}$ typically of
the order of $10^6-10^7$. 
We now compare this value with 
the average fitting parameter $\gamma_{\mathrm{reg}}=399$MHz 
of Eq.~(\ref{gamtot}) to the data of Fig.~\ref{gamvsn}, which 
is related
to the average intrinsic $Q_{\mathrm{reg}}$ factor of the WGMs as
$Q_{\mathrm{reg}}=2\pi\nu/\gamma_{\mathrm{reg}}\simeq 5\cdot10^6$,
consistently with the outcome of the fiber-taper experiment.


Thus, the overall enlargement of the average linewidths with the number of observed regular
modes 
supports
the approximations leading to Eq.~(\ref{nregmodes}).

\section{Conclusion and Discussion}
\label{concl}

We count statistics of chaotic resonances in a deformed optical microcavity
by the sole experimental detection of high$-Q$ regular modes, using the coupling
between regular and chaotic modes, which occurs via dynamical tunneling.

Being a priori unaware of the typical escape time of the chaotic modes that
effectively contribute to the excitation of the regular modes, we use
the experimental data to validate: $i)$ an entirely classical model, $ii)$ a RMT-based,
purely statistical prediction, which is independent of system-specific properties,
and, finally, $iii)$ a semiclassical correction to $ii)$, which does depend on the Lyapunov
exponent of the chaotic dynamics.
We find theory $iii)$ in the best
agreement with
the observations, particularly when a microcavity of lower deformation factors is
coupled with visible light, while prediction $ii)$ also proves adequate when working in the
infrared.

The estimation of the Ehrenfest time of quantum-to-classical correspondence
from the experimental parameters plays a key role in
framing the time scale of the decay (or typical linewidth) of the chaotic states (resonances).
The fastest escape  occurs
around Ehrenfest time, and generally within the average time of transition
of the decay of correlations from exponential to algebraic, so that the classical
description of the dynamics as fully chaotic seems appropriate.

On the other hand,
accounting for the long-lived chaotic resonances 
does not seem to be as straightforward.
Specifically, the effects of partial transport barriers,
   as well as the `stickiness'
along KAM tori and stability islands
are relegated to the fitting parameters
in the current model.
The correct detection and modelling of long-lived resonances are
therefore primary issues to be addressed by future work, especially
in perspective of a test of fractal Weyl law at optical frequencies.
Other challenges include the possibility of estimating and measuring
the amplitude of the
regular-to-chaotic mode coupling, 
as well as developing a more refined prediction for the excitation of the
regular modes.


\section{Acknowledgments}
This project was supported by the Ministry of Science and Technology of China (Grants No. 2016YFA0301302,
No. 2013CB921904, and No. 2013CB328704), the NSFC (Grants No. 61435001, No. 11654003, and No. 11474011). 

\appendix

\section{Rescaling the Lyapunov exponent}
\label{applyap}
The classical estimate of the prefactor $MN^{-1/\hat{\mu}\tau_d}$ involves Ehrenfest time,
defined for \textit{open} systems as~\cite{SchomJacq}
\begin{equation}
\tau_{Ehr} = \frac{1}{\mu}\log\frac{\tau_H}{\tau_d}.
\label{Ehr_def}
\end{equation}
Here $\mu$ is the Lyapunov exponent of the closed system,
$\tau_d$ is the mean dwell time, 
 while
$\tau_H$ is the Heisenberg time
\begin{equation}
\tau_H = \frac{h}{\Delta E},
\label{Heis_t}
\end{equation}
with $\Delta E$ mean level spacing, that is average distance (difference) bewteen consecutive
energy levels. We know, on the other hand, that $E=h\nu$, and we may therefore express
Heisenberg time in terms of the frequency spacing
\begin{equation}
\tau_H = \frac{1}{\Delta\nu},
\end{equation}
and Ehrenfest time as
\begin{equation}
\tau_{Ehr} = \frac{1}{\mu}\log\frac{N}{\Delta\Upsilon}.
\end{equation}
Here $N$ is the number of open channels as we know, whereas $\Delta\Upsilon=MT\Delta\nu$,
that is the mean frequency spacing times the Poincar\'e time (to make it dimensionless),
times the number of states $M$. In plain words, $\Delta\Upsilon$ is the frequency range of
our modes in units of the Poincar\'e time. At this point we can still write
\begin{equation}
\tau_{Ehr} = \frac{1}{\hat{\mu}}\log N
\end{equation}
as in Sec.~\ref{statchastat}, provided that
\begin{equation}
\hat{\mu} = \frac{\log N}{\log N - \log\Delta\Upsilon}\mu.
\label{resc_mu}
\end{equation}
Thus we have determined the rescaling to the Lyapunov exponent, following the definition of the
Ehrenfest time.



\section{Estimation of Ehrenfest time}
\label{app:Ehr}

Let us start from the definition of the Ehrenfest time
 \begin{equation}
 \tau_{Ehr} = \frac{1}{\mu}\ln\frac{\tau_H}{\tau_d},
  \label{tehr}
  \end{equation}
  with
  \begin{eqnarray}
 \tau_H = \frac{h}{\Delta E}  =\frac{1}{\Delta\nu}, \\
 \tau_d = \frac{M}{N}T,
 \label{tHtd}
 \end{eqnarray}
 and $T$ is the Poincar\'e time. For visible light, we observe about $100$ WGMs in a range
 of wavelengths of about
 $10$nm, hence  an estimated mean
 spacing between consecutive regular modes
 $\Delta\lambda\sim10^{-10}$. Then
 \begin{equation}
 \Delta\nu \simeq \frac{c\Delta\lambda}{\lambda^2} \sim
 \frac{2\cdot10^8\cdot10^{-10}}{6^2\cdot10^{-14}}  =
 5\cdot10^{10} \mathrm{Hz},
\label{delnu}
 \end{equation}
while
\begin{equation}
T \simeq \frac{a}{2c} \sim \frac{6\cdot10^{5}}{2\cdot10^8} =
3\cdot10^{-13} \mathrm{s},
\label{PoincT}
\end{equation}
with $a$ principal diameter of the microcavity. That way,
\begin{equation}
\tau_{Ehr} = \frac{1}{\mu}\ln\frac{\xi}{\Delta\nu AT} ,
\label{est_ther}
\end{equation}
where we took $\xi\simeq\frac{A}{\tau_d}$ (neglecting refraction into air), and
$A=2\pi(\sin\theta_{max}-\sin\theta_{th})$, area of the phase space available
to chaotic states above  $\sin\theta_{th}$.

\end{document}